\documentclass[a4paper,11pt]{article}
\usepackage{amsmath,amssymb}
\usepackage[dvipdfmx]{graphicx}
\usepackage{bm}
\usepackage{mathptmx}
\usepackage{subcaption}
\usepackage{braket}

\usepackage{pos}
\usepackage{hyperref}

\title{Phase structure analysis of CP(1) model with $\theta$ term by tensor renormalization group}

\author*[a]{Hayato Aizawa}
\author[a]{Shinji Takeda}
\author[a]{Yusuke Yoshimura}

\affiliation[a]{Institute for Theoretical Physics, Kanazawa University, Kanazawa 920-1192, Japan}

\emailAdd{h\_aizawa@hep.s.kanazawa-u.ac.jp}
\emailAdd{takeda@hep.s.kanazawa-u.ac.jp}

\abstract{We analyze the phase structure of 2d lattice CP(1) model with $\theta$ term by using the bond-weighted tensor renormalization group method. 
We propose a new tensor network representation for the model using the quadrature scheme and confirm that its accuracy is better than that of the conventional character-like expansion.
As a probe to study the phase structure, we adopt the central charge and the scaling dimensions.
The numerical results indicate an existence of critical point at $\theta=\pi$, which is consistent with the Haldane's conjecture.}

\FullConference{The 41st International Symposium on Lattice Field Theory (LATTICE2024)\\
 28 July - 3 August 2024\\
Liverpool, UK\\}

\begin{document}
\maketitle

\section{Introduction}
The two-dimensional CP(1) model  has been studied as a toy model of QCD since they share many common properties, e.g. asymptotic freedom, confinement, and $\theta$ term.
In particular, the $\theta$ term makes it difficult to use the Monte Carlo method due to the sign problem.
Thus, the CP(1) with $\theta$ term is  of interest  as a test to examine the non-perturbative phenomena including $\theta$ term.

For a given lattice field theory, one of the important task is to study a phase structure in its bare parameter space.
The 2d lattice CP(1) with $\theta$ term has two bare parameters, $\beta$ (inverse coupling) and $\theta$, and its phase structure is partially well investigated.
In the strong coupling limit at $\theta=\pi$,
it was known that the first order phase transition occurs \cite{PhysRevLett.53.637}.
Furthermore, the strong coupling expansion analysis \cite{PhysRevD.55.3966} showed that the first order transition
persists even for introducing finite $\beta$.
On the other hand, using the equivalence of O(3) and CP(1) model and the Haldane's conjecture \cite{HALDANE1983464,PhysRevLett.50.1153}, one expects that the first order phase transition turns out to be second order when increasing $\beta$ at fixed $\theta=\pi$.
In fact, the Haldane's conjecture was verified for 2d O(3) model in analytical \cite{PhysRevB.36.5291,PhysRevLett.66.2429} and numerical \cite{PhysRevLett.75.4524,PhysRevD.86.096009,PhysRevD.77.056008} studies and the universality class is shown to be the SU(2)$_{k=1}$ WZW model.
Therefore, it is likely that 2d lattice CP(1) model also have a criticality at finite $\beta$ and it belongs to the same universality class.

On the other hand, numerical phase structure analysis for 2d lattice CP(1) model with $\theta$ term is unclear so far.
A Monte Carlo study \cite{PhysRevLett.98.257203} found a critical point around $\beta=0.5$, but its resulting critical exponent is different from that of SU(2)$_{k=1}$ WZW model.
It might be attributed from the sign problem and to make clear this point one needs sign-problem-free method, like the tensor network method \cite{PhysRevLett.99.120601}.
Nevertheless, the tensor network results are also contradictory; 
first study \cite{Kawauchi:2017dnj} found a critical region starting from $\beta\approx0.4$ while
another study \cite{PhysRevD.105.054507} using different coarse-graining algorithm could not detect a criticality up to $\beta=1.1$.

In this study we make two improvements for the tensor network study of the CP(1) model: using a better initial tensor as well as CFT information to diagnose the phase structure.
As a result, we find a critical point associated with the SU(2)$_{k=1}$ WZW model as expected by the Haldane's conjecture.

\section{CP(1) model on 2d lattice}
The action of the CP(1) model with the $\theta$ term on the square lattice ($x\in\mathbb{Z}^2$) is given by
\begin{equation}
    S=-2\beta \sum_{x,\mu}
    \left[
    z^\dag(x)z(x+\hat\mu)e^{iA_\mu(x)}
    +
    z^\dag(x+\hat\mu)z(x)e^{-iA_\mu(x)}
    \right]
    -
    \frac{\theta}{2\pi}
    \sum_x \log U_p(x),
\end{equation}
where $\hat\mu$ ($\mu=0,1$) is the unit vector for $\mu$-direction.
The periodic boundary condition is imposed on both directions.
Here, $z(x)$ is a two-component complex scalar field
\begin{equation}
    z(x)=
    \left(
    \begin{array}{c}
    z_1(x)
    \\
    z_2(x)
    \end{array}
    \right)
    \in\mathbb{C}^2,
\end{equation}
and is constrained
\begin{equation}
    |z(x)|^2=1.
\end{equation}
By using U(1) gauge field $A_\mu(x)$,
the plaquette loop $U_p(x)$ is defined by
\begin{equation}
    U_p(x)=e^{i[
A_0(x)+A_1(x+\hat0)-A_0(x+\hat1)-A_1(x)
    ]}.
\end{equation}

The associated partition function is given by 
\begin{equation}
    Z(\beta,\theta)
    =\left(\prod_x\int dz(x)\right)
    \left(\prod_{x,\mu}\int_{-\pi}^{\pi}\frac{dA_\mu(x)}{2\pi}\right)
    e^{-S}
    \label{eq:Z}
\end{equation}
where the measure for the complex scalar field $dz(x)$ is given by
\begin{equation}
    \int dz(x)=C_z\int_{\mathbb{C}^2} d\mathrm{Re}(z_1(x))d\mathrm{Im}(z_1(x))d\mathrm{Re}(z_2(x))d\mathrm{Im}(z_2(x))\delta(|z(x)|^2-1).
\end{equation}
Here the normalization factor $C_z$ is chosen such that 
$   \int dz(x)=1$.

For later convenience, we define $H$, $Q$ 
\begin{equation}
    H_{zz^\prime A}=\exp\left(2\beta
     \left[
    z^\dag z^\prime e^{iA}
    +
    z^{\prime\dag}ze^{-iA}
    \right]
    \right),
    \label{eq:H}
\end{equation}
\begin{eqnarray}
    Q_{AA^\prime A^{\prime\prime}A^{\prime\prime\prime}}&=&\exp\left(\frac{\theta}{2\pi}\log \left[
    e^{i(A+A^\prime-A^{\prime\prime}-A^{\prime\prime\prime})}
    \right]\right).
    \label{eq:Q}
\end{eqnarray}
Then the partition function can be rewritten by using $H$, $Q$
\begin{equation}
    Z(\beta,\theta)
    =\left(\prod_x\int dz(x)\right)
    \left(\prod_{x,\mu}\int_{-\pi}^{\pi}\frac{dA_\mu(x)}{2\pi}\right)
    \prod_{x,\mu}H_{z(x),z(x+\hat\mu),A_\mu(x)}
    \prod_{x}Q_{A_0(x),A_1(x+\hat0),A_0(x+\hat1),A_1(x)}.
    \label{eq:ZHQ}
\end{equation}

\section{Initial tensor}
In the previous studies \cite{Kawauchi:2017dnj,PhysRevD.105.054507}, the initial tensor was made by truncating the character-like expansion \cite{PhysRevD.55.3966}, but
it is known that the expansion for the $\theta$ term has very poor convergence.
The resulting initial tensor is hence expected to be strongly affected by large truncation error,
and furthermore the truncating error was not so seriously estimated.

In our study we make a new initial tensor using the quadrature method \cite{Kadoh:2018hqq,GENZ2003187}
which approximates the integral by a finite sum.
\begin{equation}
    \int dz f(z) \approx \sum_{i=1}^{N_z} W_{i}^{(z)} f(z_i),
    \label{eq:Nz}
\end{equation}
\begin{equation}
    \int_{-\pi}^{\pi} \frac{dA}{2\pi} g(A) \approx \sum_{a=1}^{N_A} W_{a}^{(A)} g(A_a),
    \label{eq:Na}
\end{equation}
where $z_i$ ($A_a$) is a sample point, $N_z$ ($N_A$) is the number of sample points, and $W_i^{(z)}$ ($W_a^{(A)}$) is a weight of the associated quadrature for the complex scalar fields (the gauge field).
These expansions allow us to represent $Z$ as a tensor network,
\begin{equation}
    Z\approx\left(\prod_x\sum_{i_x}^{N_z}\sum_{{i_x}^\prime}^{N_z}\right)\left(\prod_{x,\mu}\sum_{a_{x,\mu}}^{N_A}\right)
    \prod_x T_{(i_{x+\hat{0}},a_{x+\hat{0},1})(i_{x+\hat{1}},a_{x+\hat{1},0})(i_x,a_{x,1})({i_x}^\prime,a_{x,0})}
    \label{eqn:tensor_rep}
\end{equation}
where the initial tensor $T$ is defined as follows
\begin{equation}
    T_{(z_1A_1)(z_2A_2)(z_3A_3)(z_4A_4)}
    =
    W_{z_4}^{(z)}
    \sqrt{W_{A_1}^{(A)}
    W_{A_2}^{(A)}
    W_{A_3}^{(A)}
    W_{A_4}^{(A)}}
    \delta_{z_3,z_4}
    H_{z_3,z_1,A_4}
    H_{z_4,z_2,A_3}
    Q_{A_4,A_1,A_2,A_3}.
    \label{eqn:tensor_naive}
\end{equation}
Note that the integers of the sum in eq.(\ref{eq:Nz}) and (\ref{eq:Na}) now turn out to be an index of the initial tensor.
One of the index of the initial tensor, say $(z_1A_1)$, has a dimension $N_z \times N_A$
which usually requires large $O(10^4)$ to keep better accuracy. 
For an actual calculation, we thus should reduce the dimension by compressing the initial tensor.
For that purpose,  we use HOTRG method \cite{PhysRevB.86.045139}
and make a projector $P^{(1)}$, which reduces the dimension from $N_z \times N_A$ to $D_{\rm c}\sim O(10-100)$,
by minimizing a cost function as follows,
\begin{equation}
    \left|T_{(z_1A_1)(z_2A_2)(z_3A_3)(z_4A_4)}-\sum_{z,A,i}T_{(zA)(z_2A_2)(z_3A_3)(z_4A_4)}P^{(1)}_{zA,i}P^{(1)\dag}_{z_1A_1,i}\right|^2.
\end{equation}
In the end, we obtain a compressed initial tensor, whose bond dimension is $D_{\rm c}$, as follows,
\begin{equation}
    T_{i_1i_2i_3i_4}^{({\rm compressed})}=\sum_{z_1,z_2,z_3,z_4,A_1,A_2,A_3,A_4}T_{(z_1A_1)(z_2A_2)(z_3A_3)(z_4A_4)}
    P^{(1)}_{z_1A_1,i_1}P^{(2)}_{z_2A_2,i_2}P^{(1)\dag}_{z_3A_3,i_3}P^{(2)\dag}_{z_4A_4,i_4}.
    \label{eqn:tensor}
\end{equation}

\begin{figure}[t]
    \centering
    \includegraphics[width=0.7\columnwidth]{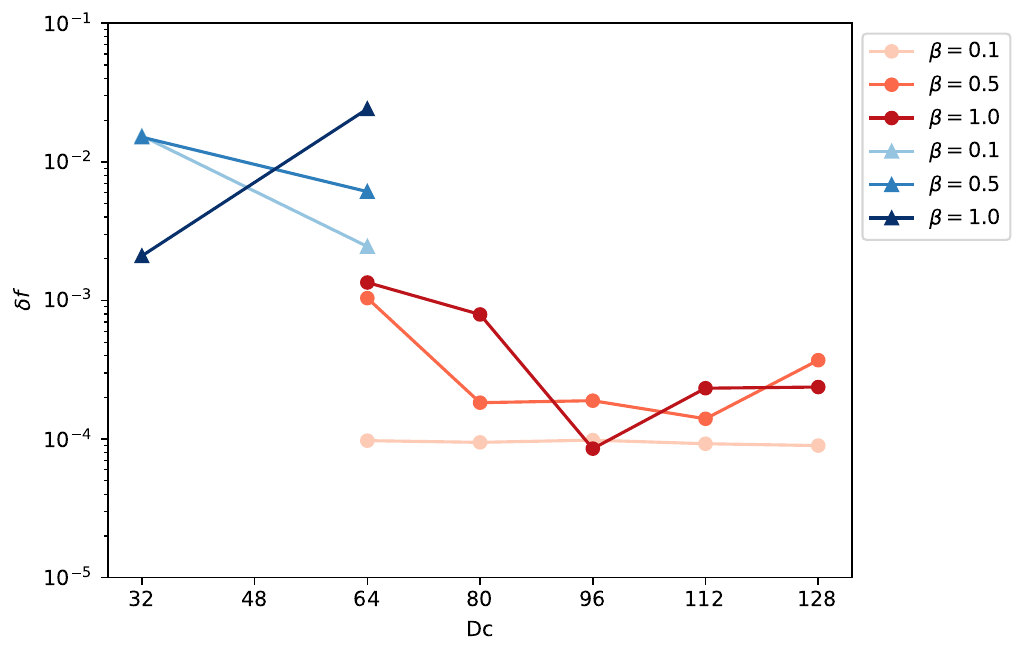}
    \caption{
The relative error of the free energy in eq.(\ref{eqn:error}) on $2\times2$ lattice as a function of $D_{\rm c}$ with
$\theta=\pi$ and $\beta=0.1-1$.
The red circles are for our new initial tensor while
the blue triangles are for the conventional character expansion.
}
    \label{fig:cp12by2_Dc_df}
\end{figure}

In order to evaluate the truncation error of the initial tensor,
we compute the free energy on $2\times2$ lattice using the initial tensor in eq.(\ref{eqn:tensor}) and compare it with the exact result.
The relative error of the free energy is defined as
\begin{equation}
    \delta f=\frac{|f_{\rm tensor}-f_{\rm exact}|}{|f_{\rm exact}|},
\label{eqn:error}
\end{equation}
where $f_{\rm exact}$ is an exact free energy which can be numerically obtained in a brute-force way for such a small lattice, and
$f_{\rm tensor}$ is obtained by exactly contracting the four initial tensors without coarse-graining step.
Figure \ref{fig:cp12by2_Dc_df} shows the relative error at $\theta=\pi$ for our new initial tensor as well as a conventional initial tensor using the character expansion.
As a result, we observe that the error of the new initial tensor is smaller than the conventional one in the range we investigated $\beta=0.1-1$.
Therefore we exclusively use the new initial tensor in the following numerical calculations.

\section{Numerical results}
In order to elucidate the phase structure, here we utilize the central charge and the scaling dimensions
which can detect a criticality and are useful to identify its universality class.
The first step to obtain the CFT information is to coarse-grain the tensor network
whose initial value is given in eq.(\ref{eqn:tensor}).
We use the bond-weighted TRG algorithm \cite{PhysRevB.105.L060402} with $k=-\frac{1}{2}$ for the coarse-graining
and then after $n$ steps a coarse-grained tensor $T^{(n)}$ is obtained.
By taking a trace for the spatial direction, one obtains a transfer matrix $M$ for system size $V=L^2=4^{n}$,
\begin{equation}
    M_{ij}=\sum_{k}T_{kikj}^{(n)}.
\end{equation}
The eigenvalues of the transfer matrix $\lambda_i$ ($i=0,1,2,\cdots$) are related with the energy spectrum $E_i$
\begin{equation}
    \lambda_i=e^{-E_iL}.
\end{equation}
Once an invariant tensor is obtained after some coarse-graining steps,
a corresponding transfer matrix, which is made from the invariant tensor, is regarded as a CFT transfer matrix.
Therefore, $\lambda_i$, which are calculated from the invariant tensor,
may be related to the central charge $c$ and the scaling dimensions $x_i$ ($i=1,2,3,\cdots$) as follows \cite{PhysRevB.80.155131}
\begin{equation}
    c=\frac{6}{\pi}\log(\lambda_0),
    \hspace{10mm}
    x_i=\frac{1}{2\pi}\log\left(\frac{\lambda_0}{\lambda_i}\right).
\end{equation}

\begin{figure}[t]
    \centering
    \includegraphics[width=0.6\columnwidth]{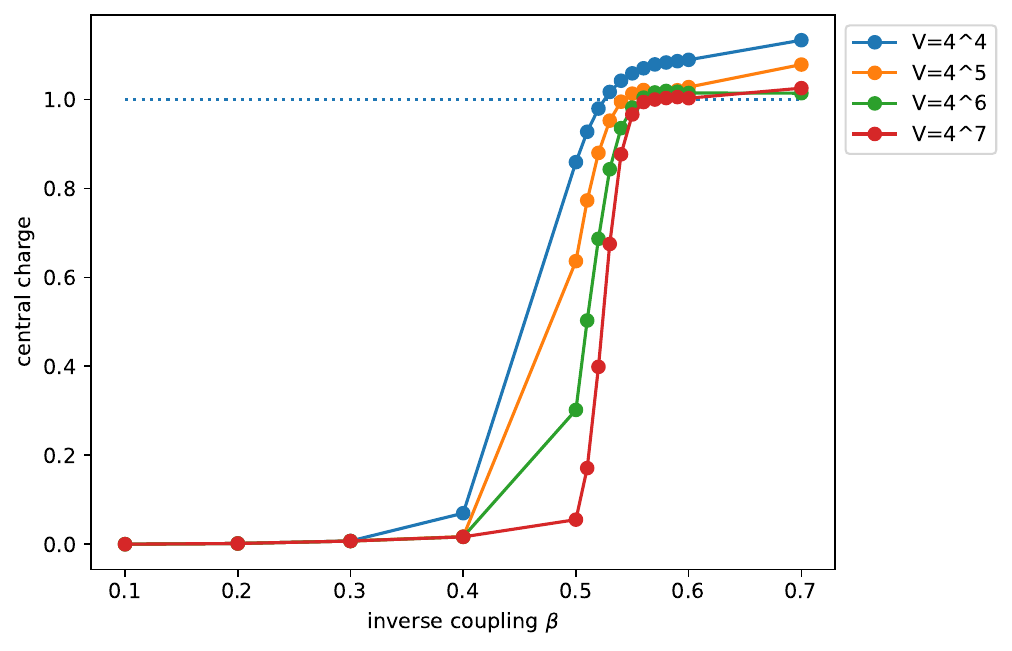}
    \caption{$\beta$-dependence of the central charge at $\theta=\pi$ for various volumes.
    The parameters of the initial tensor are $N_z=226$, $N_a=120$, and $D_{\rm c}=128$.
    For coarse-graining step, we use bond-weighted TRG $k=-\frac{1}{2}$ with the bond dimension $D_{\rm c}=128$.
}
    \label{fig:cp1_th1_beta_cc}
\end{figure}

Numerical results of the central charge as a function of $\beta$ at fixed $\theta=\pi$ is shown in Fig.~\ref{fig:cp1_th1_beta_cc}.
The parameters of the algorithm are summarized in the caption of the figure.
We observe that the central charge suddenly takes a value from zero to unity around $\beta \sim 0.55$
and the trend becomes more obvious for larger system size.
Therefore we can see that $\beta\gtrsim0.55$ is a critical region and it belongs to the free boson universality class.
\begin{figure}[t]
    \centering
\begin{minipage}{0.45\columnwidth}
    \centering
    \includegraphics[width=0.9\columnwidth]{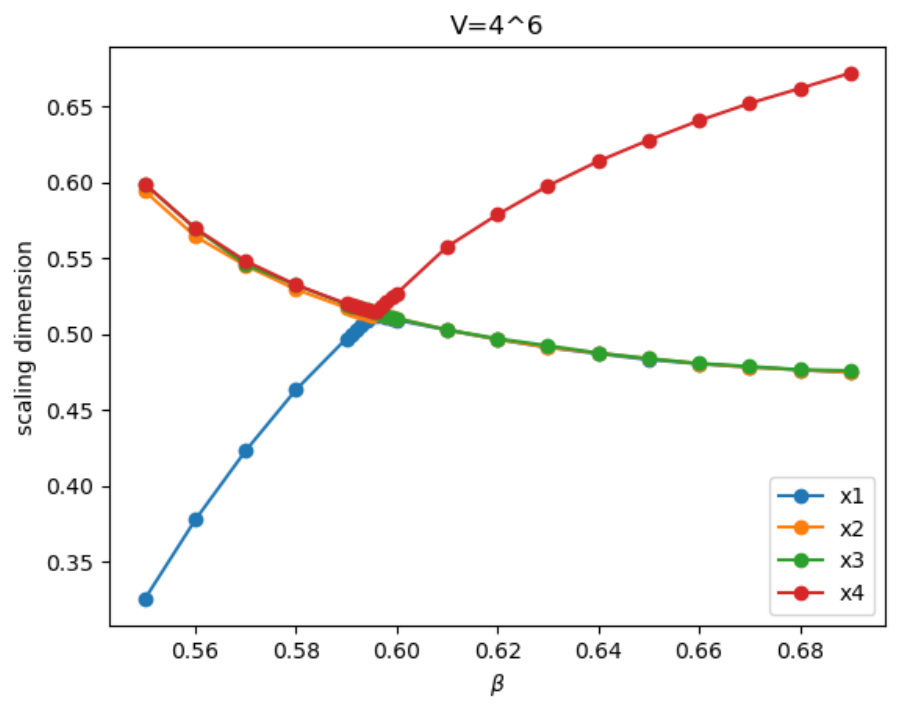}
    \label{fig:cp1-beta-sd}
\end{minipage}
\begin{minipage}{0.45\columnwidth}
    \centering
    \includegraphics[width=0.9\columnwidth]{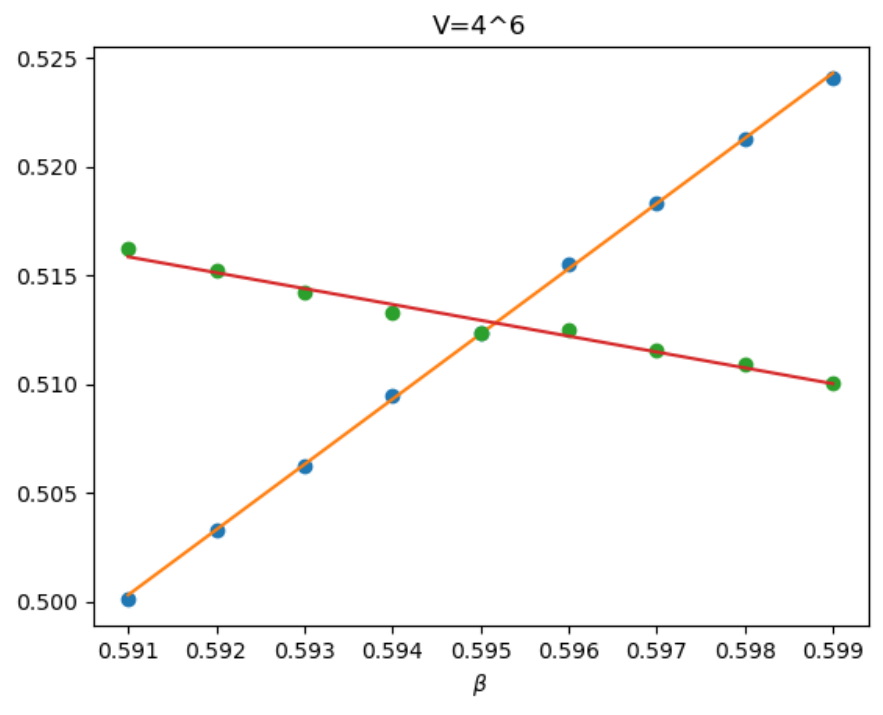}
    \label{fig:sdfit}
\end{minipage}
   \caption{
    (a): $\beta$-dependence of the first four scaling dimensions $x_i(i=1,2,3,4)$ for $V=4^6$ at $\theta=\pi$.
    (b): Zoom of the crossing region with linear fitting functions. At the crossing point ($\beta\approx0.595$), the degeneracy turns out to be quartet and it corresponds to a critical point with the universality class of SU(2)$_{k=1}$ WZW model.}
   \label{fig:sd}
\end{figure}

Figure \ref{fig:sd} shows first four scaling dimensions $x_i(i=1,2,3,4)$
at $\theta=\pi$ with the same algorithm parameter as Fig.~\ref{fig:cp1_th1_beta_cc}.
In low $\beta$ region, the lowest scaling dimension is a singlet and the following three scaling dimensions are degenerate,
while in high $\beta$ region, the lowest scaling dimension is triple degenerate and the fourth one is a singlet.
Moreover, we can see a crossing point which is roughly estimated as $\beta \approx 0.595$.
At the crossing point, the lowest scaling dimensions are 4-fold degenerate which indicates that CFT is SU(2)$_{k=1}$ WZW model.

\section{Conclusion}
We have investigated the phase structure of 2d lattice CP(1) model with $\theta$ term
by using the bond-weighted TRG algorithm and found
that along the $\theta=\pi$ line there is a critical region. 
In the study, we have made two improvements compared with the previous studies; we adopt
the quadrature scheme to make an initial tensor and the CFT-based analysis method to diagnose the phase structure.
We have confirmed that the new initial tensor is more accurate compared with the conventional one.
Furthermore, thanks to the CFT information, that is, the central charge and the scaling dimensions,
we identify the critical region and
estimate the crossing point $\beta\approx0.595$ where SU(2)$_{k=1}$ WZW CFT is realized.
This is consistent with a prediction of the Haldane's Conjecture.
\acknowledgments
This work was supported by JST SPRING, Grant Number JPMJSP2135.
\bibliography{pos_lattice2024}
\bibliographystyle{JHEP}
\end{document}